\documentclass[prb,amssymb,superscriptaddress]{revtex4}
\usepackage{graphicx}
\usepackage{bm}
\usepackage{hyperref}
\usepackage{float}
\usepackage{dcolumn}
\usepackage{color}
\usepackage{xcolor}
\usepackage{amsmath}
\usepackage{amsfonts}
\usepackage{amssymb}
\renewcommand{\arraystretch}{1.2}
\renewcommand{\text}{\mbox}

\makeatletter
\renewcommand*\env@matrix[1][\arraystretch]{%
  \edef\arraystretch{#1}%
  \hskip -\arraycolsep
  \let\@ifnextchar\new@ifnextchar
  \array{*\c@MaxMatrixCols c}}
\makeatother

\begin{document}

\title{Heavy Dirac fermions in a graphene/topological insulator hetero-junction}
\author{Wendong Cao}
\affiliation{State Key Laboratory of Low-Dimensional Quantum Physics, Department of Physics, Tsinghua University, Beijing 100084, China}

\author{Rui-Xing Zhang} 
\affiliation{Department of Physics, The Pennsylvania State University, University Park, Pennsylvania 16802-6300, USA}

\author{Peizhe Tang}
\affiliation{Department of Physics, McCullough Building, Stanford University, Stanford, California 94305-4045, USA}

\author{Gang Yang} 
\affiliation{Department of Physics, The Pennsylvania State University, University Park, Pennsylvania 16802-6300, USA}

\author{Jorge Sofo} 
\affiliation{Department of Physics, The Pennsylvania State University, University Park, Pennsylvania 16802-6300, USA}

\author{Wenhui Duan}
\affiliation{State Key Laboratory of Low-Dimensional Quantum Physics, Department of Physics, Tsinghua University, Beijing 100084, China}
\affiliation{Institute for Advanced Study, Tsinghua University, Beijing 100084, China}
\affiliation{Collaborative Innovation Center of Quantum Matter, Beijing 100084, China}

\author{Chao-Xing Liu}
\email{e-mail:  cxl56@psu.edu}
\affiliation{Department of Physics, The Pennsylvania State University, University Park, Pennsylvania 16802-6300, USA}

\date{\today}

\begin{abstract}
The low energy physics of both graphene and surface states of three-dimensional topological insulators is described by gapless Dirac fermions with linear dispersion. In this work, we predict the emergence of a ``heavy" Dirac fermion in a graphene/topological insulator hetero-junction, where the linear term almost vanishes and the corresponding energy dispersion becomes highly non-linear. By combining {\it ab initio} calculations and an effective low-energy model, we show explicitly how strong hybridization between Dirac fermions in graphene and the surface states of topological insulators can reduce the Fermi velocity of Dirac fermions. Due to the negligible linear term, interaction effects will be greatly enhanced and can drive ``heavy'' Dirac fermion states into the half quantum Hall state with non-zero Hall conductance.
\end{abstract}

\maketitle

\section{Introduction}
Two-dimensional Dirac physics has aroused great interest in condensed matter physics ever since the discovery of graphene \cite{zhang2005experimental,novoselov2005two,neto2009electronic} and topological insulators (TIs)\cite{RevModPhys.82.3045,RevModPhys.83.1057,moore2010birth} due to its importance in both fundamental physics and device applications. In graphene, gapless Dirac cones exist at the momenta $K$ and $K'$ point in the Brillion zone (BZ) with a large Fermi velocity $\simeq$ $1\times10^6$ m/s, which is about 1/300 of the speed of light\cite{PhysRev.71.622}, and results in a high mobility for electron transport in graphene. Due to its high mobility, graphene is believed to possess the potential in the applications involving fast speed electronic devices \cite{nl102824h,RevModPhys.83.407,novoselov2012roadmap}. On the other hand, when a gap (mass) is opened for Dirac cones, the so-called ``parity anomaly'', which was first known in high energy physics\cite{PhysRevD.29.2375},  can occur \cite{PhysRevLett.61.2015} and lead to a large variety of topological states, including the quantum anomalous Hall (QAH) effect \cite{PhysRevLett.61.2015} and the quantum spin Hall (QSH) effect \cite{PhysRevLett.95.226801}, in graphene. Furthermore, it was predicted that these topological states in graphene can be spontaneously induced by interaction, leading to the so-called ``topological Mott insulators''\cite{PhysRevLett.100.156401}. However, due to the large Fermi velocity, the density of states vanishes rapidly near the Dirac cone, and the critical interaction strength is relatively large. Thus, the reduction of Fermi velocity in Dirac fermions (or equivalently a ``heavy'' Dirac fermion) is valuable for exploring new topological states in the Dirac systems.

 In this work, we explore Dirac physics in a hetero-junction with graphene on top of a TI. The QSH effect with a large energy gap has been predicted for graphene coupled to TI thin films\cite{doi:10.1021/nl4037214,PhysRevB.87.075442}. Here we focus on the case of graphene coupled to a {\it single} surface state of a three dimensional TI, which can be realized in experiments when the thickness of the TI is large enough so that the topological surface states on opposite surfaces are decoupled. In this case, we find that due to the strong hybridization between the four Dirac cones from graphene (two valleys and two spins) and one Dirac cone from the topological surface state, four of the resulting states will be gapped while one remains gapless with the Fermi velocity significantly reduced. By constructing an effective model of this system based on first principles calculations, we explore the underlying physical reason why the linear dispersion relation term is so small. As a consequence, the ``heavy'' Dirac fermion has a much larger density of states near the Dirac point, and a relatively small interaction can drive the system into a gaped topological phase with non-zero Hall conductance (parity anomaly). Below, we will first construct an effective model for this hetero-junction that reproduces the physics observed in our first-principles calculations, and then study the effects of interactions at the mean field level.

\section{Effective model for graphene/topological insulator hetero-junctions}
\begin{figure}[H]
\centering
\includegraphics[width=0.6\linewidth]{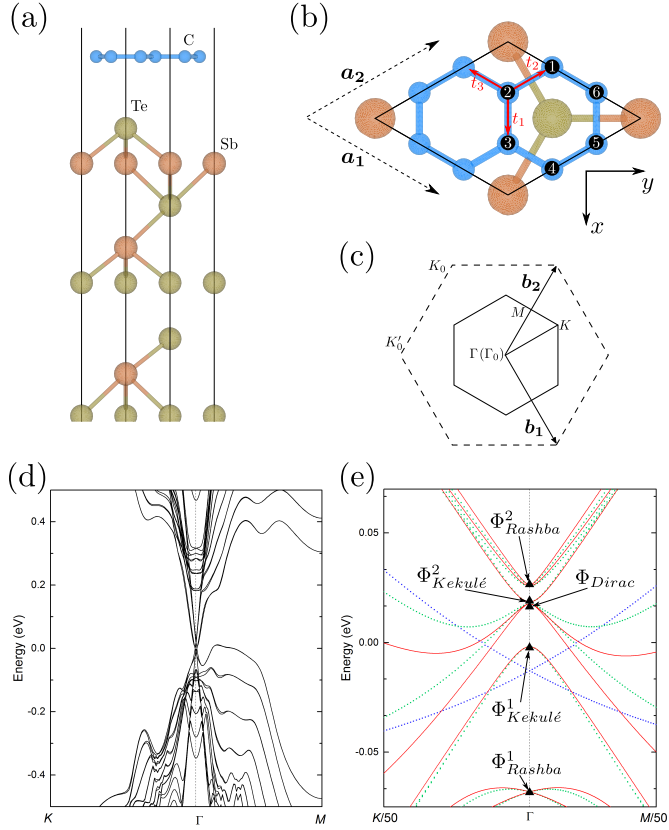}
\caption{(a) Side view of the hetero-junction with graphene and the top few atomic layers of Sb$_2$Te$_3$. (b) Top view of the hetero-junction for graphene and the two topmost atomic layers of Sb$_2$Te$_3$. The black solid rhombus is a unit cell and two in-plane lattice vectors are $\{\bm{a_1},\bm{a_2}\}$. The three red arrows label three hopping processes with the amplitudes $\{t_1,t_2,t_3\}$ between nearest carbon atoms in graphene. (c) The folded Brillouin zone of the unit cell in (b) are denoted by solid lines and high-symmetric points ($\Gamma$, $M$ and $K$) are labelled. $K_0$ and $K_0'$ are also presented in the original Brillouin zone of the pristine graphene (dashed lines). (d) Calculated band structure along the high-symmetric lines $K-\Gamma-M$. (e) Low-energy band structure around the $\Gamma$ point from \emph{ab initio} calculation (green dots) and the corresponding effective Hamiltonian $H_{full}$ in Eq.[\ref{equ:equ1}] (red lines). $\Phi_{Rashba}^1$, $\Phi_{Kekul{\acute{e}}}^1$, $\Phi_{Dirac}$, $\Phi_{Kekul{\acute{e}}}^2$, $\Phi_{Rashba}^2$ labels bands that we are interested in.  The blue dots represent the bottom surface states, which are decoupled from graphene. }
\label{Fig:fig1}
\end{figure}

Firstly, we perform {\it ab intio} calculations for a hetero-junction with graphene on a Sb$_2$Te$_3$ film with eight quintuple (QL) layers, as shown in Fig. \ref{Fig:fig1}(a). Due to the lattice mismatch between graphene and Sb$_2$Te$_3$ films (substrate), the slab model used in this calculation contains one unit cell of Sb$_2$Te$_3$ films \cite{wyckoff1960crystal}, whose in-plane lattice constant corresponds to that of the $\sqrt{3}\times\sqrt{3}$ supercell of graphene with $0.3\%$ compression. The top view of the most stable configuration is shown in Fig. \ref{Fig:fig1}(b) in which the topmost Te atom locates at the center of the hexagonal ring of graphene \cite{PhysRevB.87.075442}. From our calculations, it is found that the equilibrium distance between graphene and Sb$_2$Te$_3$ is $3.456~\rm \AA$ and the binding energy is $41.9$ meV per carbon atom. Since the $\sqrt{3}\times\sqrt{3}$ supercell of graphene is used in our simulation, the BZ of the hetero-junction is folded, and reduced to one third of the original BZ of the intrinsic graphene (the original and folded BZs are shown in Fig. \ref{Fig:fig1}(c)), and thus the $K_0$ and $K_0'$ in the original BZ are mapped into $\Gamma$ in the folded BZ. Consequently, the Dirac cones from both graphene and TI appear at the same momentum $\Gamma$, and we only need to focus on the low energy physics around the $\Gamma$ point. From the {\it ab initio} calculations, the band structure of the hetero-junction is shown in Fig. \ref{Fig:fig1}(d) and its low energy bands are zoomed in (dotted lines) in Fig. \ref{Fig:fig1}(e). Spin-orbit coupling (SOC) is included in the calculations and more details about $\emph{ab initio}$ calculation method can be found in the appendix A. In Fig. \ref{Fig:fig1}(e), via analyzing the wave functions at the $\Gamma$ point around the Fermi level, we find that the bottom surface states of TIs has no hybridization with graphene and contribute to the gapless Dirac cone labelled by blue-dotted lines in Fig. \ref{Fig:fig1}(e). The bands labelled by green-dotted lines originate from the hybridization of energy bands in graphene and top surface states of TIs. Therefore, we focus on these energy bands (marked by the green-dotted lines) and unveil underlying physics for the ``heavy" Dirac Fermion. 

To understand the hybridization between these Dirac fermions, we first neglect the topological surface states of Sb$_2$Te$_3$ films and consider only the $p_z$ orbitals of graphene under the environment of the Sb$_2$Te$_3$ substrate. For intrinsic graphene, there are two sub-lattice sites in one unit cell and the low energy physics can be effectively described by a two-dimensional Dirac type of Hamiltonian around $K_0$ and $K_0'$ in the original BZ \cite{neto2009electronic}. In contrast, due to the influence of the substrate, one unit cell of this hetero-junction contains a hexagon (benzene ring) with six equivalent carbon sites and the hopping terms between them could be described by three hopping parameters, denoted as $t_1$, $t_2$ and $t_3$ in Fig. \ref{Fig:fig1}(b). Therefore, a $6\times 6$ tight-binding model \cite{neto2009electronic} on the basis of $|p_z,n\rangle$, $n=1,\dots,6$ is used to describe this system (see Eq. [\ref{equ:tbspinless}] in the appendix B) in which, for the basis $|p_z,n\rangle$, $p_z$ is the atomic orbital and $n$ represents the site index.

For convenience, we perform a unitary transformation to change the basis into eigenstates of the rotation operator. We consider the six-fold rotation operation $C_6$ and the corresponding rotation symmetric basis are denoted as $C_6|L'_z\rangle=exp(-L'_z\frac{i\pi}{3})|L'_z\rangle$, where $L'_z=\pm 2, \pm 1, 0, 3$ labels six eigen states for $C_6$ rotation. (The detailed form of unitary transformation, as well as the basis $|L'_z\rangle$, is given by Eq. [\ref{equ:6x6trans}] in the appendix B and the form of the effective Hamiltonian expanded around ${\bf k}=0$ is given by Eq. [\ref{equ:6x6kpmodel}] in the appendix B). We emphasize that six-fold rotation symemtry only exists in pristine graphene ($t_1=t_2=t_3$), in which two gapless Dirac fermions appear under the basis $|L'_z=\pm 2\rangle$ and $|L'_z=\pm 1\rangle$ and the states labeled by $|L'_z=0\rangle$ and $|L'_z=3\rangle$ are gapped. This is because the states $|L'_z=\pm 2\rangle$ and $|L'_z=\pm 1\rangle$ originate from the states in the $K_0$ and $K_0'$ while the states $|L'_z=0\rangle$ and $|L'_z=3\rangle$ from those at $\Gamma_0$ in the original BZ. Therefore, in the following discussion, we only focus on the basis $|L'_z=\pm 2\rangle$ and $|L'_z=\pm 1\rangle$. Due to SOC, the orbital angular momentum is not a good quantum number and the total angular momentum should be considered. Thus, these eigenstates can be labeled by $|J'_z=-\frac{3}{2},\uparrow\rangle$, $|-\frac{1}{2},\uparrow\rangle, |\frac{5}{2},\uparrow\rangle, |\frac{3}{2},\uparrow\rangle, |\frac{3}{2},\downarrow\rangle, |\frac{1}{2},\downarrow\rangle, |-\frac{5}{2},\downarrow\rangle$ and $|-\frac{3}{2},\downarrow\rangle$.

Due to the hybridizations with the Sb$_2$Te$_3$ substrate, the six-fold rotation symmetry $C_6$ is broken down to three-fold rotation $C_3$, which is known as Kekul{\'{e}} modulation\cite{PhysRevLett.98.186809}. In this case, we may re-label our basis states  $|L'_z=\pm 2\rangle$ and $|L'_z=\pm 1\rangle$ by the eigenvalues of $C_3$ rotation, denoted as $|L_z,\eta\rangle$, where $C_3|L_z,\eta\rangle=exp(-L_z\frac{2\pi i}{3})|L_z,\eta\rangle$ with $L_z=\pm 1$, $\eta=\pm 1$. The index $\eta$ is introduced to distinguish two degenerate states with the same $L_z$. The explicit form of $|L_z,\eta\rangle$ in the basis set of $\{|p_z,n\rangle\}$ and more details about the additional index $\eta$ can be found in the appendix C. 
Under the basis $|+1,+1\rangle, |-1,+1\rangle, |-1,-1\rangle, |+1,-1\rangle$ (or equivalently $|L'_z=-2\rangle, |-1\rangle, |2\rangle, |1\rangle$), the effective Hamiltonian is written as
\begin{eqnarray}
	H_{G,4\times 4}=\left(\begin{array}{cccc}
-\Delta\mathrm{cos}\theta	&\hbar v_{f}^{G}{k}_{+}	&0	&\Delta\mathrm{sin}\theta	\\
\hbar v_{f}^{G}{k}_{-}	&+\Delta\mathrm{cos}\theta	&\Delta\mathrm{sin}\theta	&0	\\
0	&\Delta\mathrm{sin}\theta	&-\Delta\mathrm{cos}\theta	&-\hbar v_{f}^{G}{k}_{-}	\\
\Delta\mathrm{sin}\theta	&0	&-\hbar v_{f}^{G}{k}_{+}	&+\Delta\mathrm{cos}\theta	\\
\end{array}\right)
\label{eq:Kekule}
\end{eqnarray}
where $v_{f}^{G}$ is the Fermi velocity of graphene, $k_{\pm}=k_x\pm ik_y$, $\Delta=\sqrt{t_1^2+t_2^2+t_3^2-t_1t_2-t_2t_3-t_3t_1}$ and $\tan\theta=\sqrt{3}(t_1-t_2)/(t_1+t_2-2t_3)$. The Kekul{\'{e}} modulation terms, described by two independent parameters $\Delta$ and $\theta$, can lead to a gap openning with the size of $2\Delta$ for both Dirac cones.

When SOC is considered, the basis set $\{|L_z,\eta\rangle\}$ should be enlarged and $L_z$ is replaced by the total angular momentum along the $z$ direction ($J_z=L_z+S_z$). The new basis states for graphene take the form of $|J_z=\frac{3}{2},\eta=+1\rangle, |-\frac{1}{2},+1\rangle$, $|-\frac{1}{2},-1\rangle, |\frac{3}{2},-1\rangle, |-\frac{3}{2},-1\rangle, |\frac{1}{2},-1\rangle, |\frac{1}{2},+1\rangle$ and $|-\frac{3}{2},+1\rangle$. Given the coupling with the substrate, the full low-energy effective Hamiltonian around the $\Gamma$ point is written as
\begin{equation}
	H_{full}=\begin{pmatrix}
H_{GG}&H_{GS}\\
H_{GS}^{\dagger}&H_{SS}\\
\end{pmatrix}.
\label{equ:equ1}
\end{equation}
Here $H_{GG}$ is the Hamiltonian of graphene which takes spin into account. The detailed form of $H_{GG}$ is given by Eq. [\ref{equ:8x8kpmodel}] in the appendix B. $H_{SS}$ describes topological surface states on top surface of Sb$_2$Te$_3$ films and is given by \begin{equation}H_{SS}=\mu^S+
\begin{pmatrix}
0	&i\hbar v_f^S	k_-\\
-i\hbar v_f^S k_+	& 0\\
\end{pmatrix}
\end{equation} where $v_{f}^{S}$ is the Fermi velocity of top surface states, and $\mu^S$ denotes the corresponding chemical potential. The basis states for topological surface states can also be labelled by their eigenvalues of $J_z$, $\{|J^S_z=\pm\frac{1}{2}\rangle\}$. Furthermore, we can construct the hybridization Hamiltonian $H_{GS}$ between graphene and TIs directly from the hopping process between the $p_z$ orbitals of carbon atoms and all the $p$ orbitals of Se and Te atoms. The explicit form of $H_{GS}$ is shown in Eq. [\ref{equ:HGS}] in the appendix B.

With the effective Hamiltonian $H_{full}$ in Eq. \ref{equ:equ1}, we calculate the energy spectrum of graphene/TIs hetero-structure, as shown by red lines in Fig. \ref{Fig:fig1}(e). The corresponding parameters for $H_{full}$ are listed in the Table \ref{tab:fittingpara} of the appendix B. The effective model reproduces well the band structure from first-principles calculations (dotted-green lines), especially for the bands labeled by $\Phi_{Rashba}^1$, $\Phi_{Kekul{\acute{e}}}^1$, $\Phi_{Kekul{\acute{e}}}^2$ and $\Phi_{Rashba}^2$ in Fig. \ref{Fig:fig1}(e). The bands $\Phi_{Rashba}^1$ and $\Phi_{Rashba}^2$ mainly come from the hybridiation between the topological surface states $\{|J^S_z=\pm\frac{1}{2}\rangle\}$ and graphene states $|-\frac{1}{2},\pm 1\rangle$ and $|\frac{1}{2},\pm 1\rangle$ (or equivalently $|J'_z=-\frac{1}{2},\uparrow\rangle$ and $|J'_z=\frac{1}{2},\downarrow\rangle$). Due to strong SOC in TIs, these bands reveal strong Rashba type of spin splitting and thus are labeled by ``Rashba'' bands. Both $\Phi_{Kekul{\acute{e}}}^1$ and $\Phi_{Kekul{\acute{e}}}^2$ orginate from four graphene states $|-\frac{3}{2},\pm 1\rangle$ and $|\frac{3}{2},\pm 1\rangle$ (or equivalently $|J'_z=\pm\frac{3}{2},\uparrow(\downarrow)\rangle$). It should be noted that all these four states can hybridize with each other since the $C_6$ rotation symmetry is broken to the $C_3$ rotation symmetry due to the Kekul{\'{e}} modulation (thus dubbed as ``Kekul{\'{e}}" bands). The gap between $\Phi_{Kekul{\acute{e}}}^1$ and $\Phi_{Kekul{\acute{e}}}^2$ requires the Kekul{\'{e}} modulation and thus is smaller than that between $\Phi_{Rashba}^1$ and $\Phi_{Rashba}^2$, which only depends on the coupling strength between graphene and TIs. 

Within the gap between $\Phi_{Kekul{\acute{e}}}^1$ and $\Phi_{Kekul{\acute{e}}}^2$, there are other two bands, labeled by $\Phi_{Dirac}$, which are dominated by the states $|\frac{1}{2},+1\rangle$ and $|-\frac{1}{2},-1\rangle$ (or correspondingly $|J'_z=-\frac{5}{2},\downarrow\rangle$ and $|J'_z=\frac{5}{2},\uparrow\rangle$) . For these two bands, the effective model only recovers the energy dispersion close to the $\Gamma$ point. For larger momentum, a highly nonlinear behaviour can be observed from the first-principles calculations and suggests that higher momentum terms are dominant for these two bands. To get an effective description for these two bands, we apply the L$\ddot{o}$wdin perturbation theory to the full Hamiltonian $H_{full}$ (\ref{equ:equ1}) and project it into the low-energy subspace of $\Phi_{Dirac}$. Up to the third order in the momentum $k$, we obtain the following two-band effective Hamiltonian:
\begin{equation}
H_{eff}(k)=(C_2k^2+C_3)\hat{z}\cdot(\vec{\sigma}\times{\bf k})-\frac{C_1}{2}(k_+^3+k_-^3)\sigma_z+(e_0-C_0k^2)I_{2\times2}
\label{eq:Heff}
\end{equation}
where $e_0=0.01698$ eV, $C_0=116.1186$ eV, $C_1=32.2418$ eV, $C_2=936.4909$ eV, $C_3=0.02039$ eV are obtained by fitting to the energy dispersion. $\vec{\sigma}=(\sigma_x,\sigma_y,\sigma_z)$ are the Pauli matrices. For further discussions, we can ignore the identity terms and re-write the effective Hamiltonian in a nice form: $H_{eff}=\sum_{i=x,y,z} d_i(k)\sigma_i$ with $d_x=(C_2k^2+C_3)k_y,d_y=-(C_2k^2+C_3)k_x,d_z=-\frac{C_1}{2}(k_+^3+k_-^3)$. In Eq. \ref{eq:Heff}, the parameter $C_3$ describes the linear dispersion and corresponds to the Fermi velocity around $1.3\times10^4~ m\cdot s^{-1}$, which is greatly reduced compared to its original value $3\times10^5 ~m\cdot s^{-1}$ (the Fermi velocity of TI surface states). This estimate quantitatively reveals that the Fermi velocity of the bands $\Phi_{Dirac}$ is quite small and the dominating terms in Eq. (\ref{eq:Heff}) are the cubic terms.

The heavy Dirac fermion of $\Phi_{Dirac}$ can be physically understood from the form of their wave functions. 
One notices that the spin components of $|\frac{1}{2},+1\rangle$ and $|-\frac{1}{2},-1\rangle$ are opposite, so the coupling between them has to involve a spin-flip process. Since SOC in graphene is negligible \cite{PhysRevB.75.041401}, this coupling can only originate from the hybridization with the TI surface states. 
Therefore, the interactions between $|-\frac{1}{2},-1\rangle$ and $|\frac{1}{2},+1\rangle$ in graphene can only be mediated by the interlayer coupling between graphene and the topological surface states through second order (or higher order) perturbations. A detailed analysis of the possible coupling processes is shown in the appendix D. We conclude that the weak coupling between $|\frac{1}{2},+1\rangle$ and $|-\frac{1}{2},-1\rangle$ makes the resulting Dirac fermion much heavier than that of the TI surface states. 



\section{Interaction effect in graphene/topological insulator hetero-junctions}
Based on the above low-energy effective Hamiltonian (\ref{eq:Heff}), we will next discuss interaction effect in this system. We consider the Hubbard repulsion interaction and write the whole Hamiltonian in the real space as
\begin{equation}
H=\sum_{\langle r,r'\rangle}\Psi^{\dagger}(r) H_0(r,r') \Psi(r')+ U\sum_{r} \psi_1^{\dagger}(r) \psi_1(r) \psi_2^{\dagger}(r) \psi_2(r)
\end{equation}
where the spinor $\Psi(r)=(\psi_1(r),\psi_2(r))^T$ is written under the basis $|\frac{1}{2},+1\rangle$ and $|-\frac{1}{2},-1\rangle$. Based on the mean field approximation (details are shown in the appendix E), the above interaction Hamiltonian can be decomposed into the form 
\begin{equation}
H=\sum_k \Psi^{\dagger}(k)[H_{eff}(k)-(\Delta_y\sigma_y+\Delta_z\sigma_z)]\Psi(k)+\frac{1}{U}(\Delta_y^2+\Delta_z^2)
\end{equation}
where the order parameters $\Delta_{y,z}$ are defined as follows:
\begin{equation}
\Delta_y=\frac{U}{2}A\langle\Psi^{\dagger}\sigma_y\Psi\rangle,\ \Delta_z=\frac{U}{2}A\langle\Psi^{\dagger}\sigma_z\Psi\rangle.
\end{equation}
 $A$ is the surface area of the sample. Thus, we find two types of order parameters leading to the gap opening of heavy Dirac fermions in the hetero-structure. These order parameters $\Delta_y$ and $\Delta_z$ both break time reversal symmetry, which leads to spontaneous half quantum Hall effect of the Dirac fermions. In addition, $\Delta_y$ also breaks three-fold in-plane rotational symmetry and can be recognized as a type of nematic phases.

\begin{figure}[t]
\centering
\includegraphics[width=0.5\linewidth]{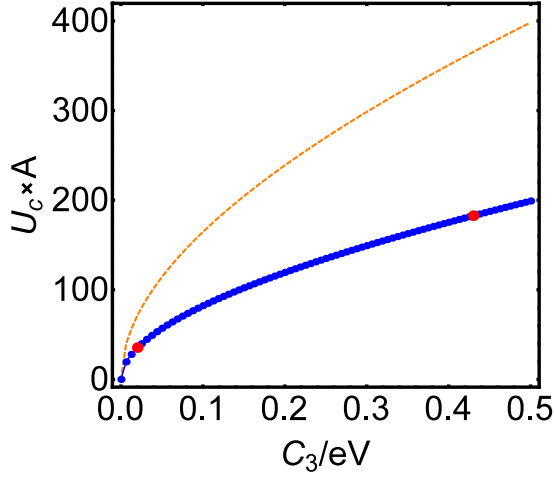}
\caption{Critical interaction strength evolving with the coefficient of linear term $C_3$ is plotted. The solid blue (orange dashed) line shows the evolution of $U_c^z$ ($U_c^y$) with $C_3$. Two red dots shows the $U_c$ required for an isolated TI surface state or a graphene/TI junction. The vertical axis is for $U_c$ times sample area, so the unit will depend on the sample size.}
\label{Fig:interaction}
\end{figure}

By minimizing the free energy, we obtain the self-consistent equations for order parameters $\Delta_i$ ($i=y,z$)
\begin{equation}
\frac{1}{U}\Delta_i=\frac{1}{2}\sum_k \frac{\Delta_i-d_i(k)}{\sqrt{d_x(k)^2+(d_y(k)-\Delta_y)^2+(d_z(k)-\Delta_z)^2}}. 
\label{eq:self-consistent eq}
\end{equation}
We emphasize that when the coefficient of linear term $C_3$ vanishes, the density of states (DOS) in the above Eq. (\ref{eq:self-consistent eq}) is divergent at the Dirac point, thus leading to an instability for heavy Dirac fermions. An infinitesimal interaction $U$ will yield a gap and drive the system into the ordered phase. With a finite but small $C_3$, the DOS vanishes at the Dirac point, and thus, a finite interaction is needed to open a gap at the Dirac point. The phase boundary is characterized by the critical interaction strength $U_c$, under which the order parameters take the limit $\lim_{U\rightarrow U_c}\Delta_i=0^+$. The analytical expression of $U_c$ is given by (see appendix F for a detailed derivation)
\begin{equation}
\frac{1}{U_c^i}=\frac{1}{2}\sum_k [\frac{1}{\sqrt{d_x(k)^2+d_y(k)^2+d_z(k)^2}}-\frac{d_i(k)^2}{[d_x(k)^2+d_y(k)^2+d_z(k)^2]^{\frac{3}{2}}}]. 
\label{Eq:Critical interaction}
\end{equation}
Here we arrived at two inequivalent self-consistent equations for $U_c^i$ ($i=y,z$) and the physical phase boundary happens for $U=\min\{U_c^y,U_c^z\}$.

The dependence of the critical interaction strength $U_c^i$ as a function of $C_3$ is shown in Fig. \ref{Fig:interaction}. We find that $U_c^z$ (plotted in a solid blue line) is always smaller than $U_c^y$ (plotted in an orange dashed line). Therefore, the blue line in Fig. \ref{Fig:interaction} determines the phase transition line in the real system. This numerical calculation also verifies our previous expectation that (i) when $C_3$ is zero, the required $U_c$ is also zero, and (ii) as $C_3$ increases, $U_c$ also increases from zero and a finite interaction strength is required to drive the system into the ordered phase.
In Fig. \ref{Fig:interaction}, we also show the $U_c$ for both an isolated TI surface state ($C_3=0.43$ eV) and a graphene/TI junction ($C_3=0.02$ eV) in red dots. 
One can see that the required critical interaction for ordered phases is almost one order of magnitude smaller in graphene/TI junctions, compared to that in pristine graphene.

\section{Discussion and Conclusion}
In this work, we demonstrate the existence of a heavy Dirac fermion with a small Fermi velocity and highly non-linear energy dispersion in the hetero-junction of graphene and a TI film, and unveil the underlying physical origins of the strong coupling between the graphene bands and the topological surface states of TI films, by combining \emph{ab initio} calculations with the low energy effective models. Due to the significant reduction of Fermi velocity, the low-energy states in graphene/TI junctions are more unstable in the presence of interactions. One may notice that the energy dispersion from first-principles calculations shows a larger non-linearity compared to that for the effective model in Fig. \ref{Fig:fig1}. This suggests that an even larger DOS appears near the Dirac cone and the corresponding critical interaction strength would be further reduced. The decreased Fermi velocity and the enhanced DOS of the Dirac fermion can be verified in experiments of angle-resolved photo-emission spectroscopy \cite{zhou2006,chen2009,xia2009} and scanning tunneling spectroscopy \cite{roushan2009,zhang2009}. If the interaction effect in this system is strong enough, topological electromagnetic effect \cite{qi2008,essin2009} can be spontaneously realized in this hetero-junction and one dimensional chiral fermions are expected to exist at the domain wall of this gaped phase. 

\section{Acknowledgement}
We would like to thank Nitin Samarth and Jun Zhu for the helpful discussions. W.C. and W.D. acknowledge the support from the National Natural Science Foundation of China (Grant No. 11334006). C.-X.L. acknowledges the support from Office of Naval Research (Grant No. N00014-15-1-2675).


\renewcommand{\theequation}{B.\arabic{equation}}
  \renewcommand{\thetable}{B.\arabic{table}}
\setcounter{equation}{0}  
\setcounter{table}{0}
\section*{Appendix}  
\subsection{Details of \emph{ab initio} calculations}
The calculations were carried out by using density-functional theory (DFT) with a plane wave basis set and the frozen projector augmented wave method for the treatment of the core electrons \cite{PhysRevB.50.17953,PhysRevB.59.1758}, as implemented in the Vienna \textit{ab initio} simulation package \cite{PhysRevB.54.11169}. A plane wave basis set with a kinetic energy cutoff of $\mathrm{450~eV}$ was used. Before the band structure calculation, graphene and the four topmost atomic layers of the substrate are fully relaxed until the residual forces are less than $1\times 10^{-3}~\mathrm{eV/\AA}$. 
During the relaxation process, the van der Waals correction of DFT-D2 method of Grimme is included \cite{grimme2006}. 
With spin-orbit coupling included, the band structure was calculated along high symmetry lines near the $\Gamma$ point.  In both geometry optimization and electronic calculation, the Monkhorst-Pack $k$ points are $15\times 15\times 1$.

\subsection{Explicit form of the effective Hamiltonian $H$}
With $p_z$ orbitals of the six carbon atoms being the basis states $\{|p_z,n\rangle,n=1,...,6\}$, the spinless tight-binding Hamiltonian of graphene is
\begin{equation}H_{tb}=
	\begin{pmatrix}
	0&	t_2&	0&	t_3e^{i\bm{k}\cdot(\bm{a_2}-\bm{a_1})}&	0&	t_1\\
	&	0&	t_1&	0&	t_3e^{-i\bm{k}\cdot\bm{a_1}}&	0\\
	&	&	0&	t_2&	0&	t_3e^{-i\bm{k}\cdot\bm{a_2}}\\
	&	&	&	0&	t_1&	0\\
	&	&h.c.&	&	0&	t_2\\
	&	&	&	&	&	0\\
	\end{pmatrix}, 
	\label{equ:tbspinless}
\end{equation} where the index $n$ for carbon atoms can be seen in Fig. \ref{Fig:fig1}. To reveal the low-energy Dirac physics, we transform the basis states to $\{|L_z'=\pm 2,\pm 1,0,3\rangle\}$, where $C_6|L_z'\rangle=\omega^{-2L_z'}|L_z'\rangle$, $C_6$ is the six-fold rotation operator and $\omega=exp(\frac{i\pi}{6})$. The basis transformation matrix $U_{6\times 6}$ is
\begin{equation}
\begin{pmatrix}
|-2\rangle&|-1\rangle&|0\rangle&|1\rangle&|2\rangle&|3\rangle\\
\hline
1&\omega^{-3}&1&\omega^3&1&1\\
\omega^{-4}&\omega^{-5}&1&\omega^5&\omega^{4}&-1\\
\omega^{4}&\omega^{5}&1&\omega^{-5}&\omega^{-4}&1\\
1&\omega^{3}&1&\omega^{-3}&1&-1\\
\omega^{-4}&\omega&1&\omega^{-1}&\omega^{4}&1\\
\omega^{4}&\omega^{-1}&1&\omega&\omega^{-4}&-1\\
\end{pmatrix}. 
\label{equ:6x6trans}
\end{equation}
After the basis transformation, we can expand the spinless Hamiltonian around the $\Gamma$ point
\begin{equation}
\begin{pmatrix}
|-2\rangle&|-1\rangle&|0\rangle&|1\rangle&|2\rangle&|3\rangle\\
\hline
-\Delta\mathrm{cos}\theta	&\hbar v_{f}^{G}{k}_{+}&0&0&\Delta\mathrm{sin}\theta&0\\
\hbar v_{f}^{G}{k}_{-}	&+\Delta\mathrm{cos}\theta	&0&\Delta\mathrm{sin}\theta&0&0\\
0&0&t_1+t_2+t_3&0&0&0\\
0&\Delta\mathrm{sin}\theta&0&-\Delta\mathrm{cos}\theta	&-\hbar v_{f}^{G}{k}_{-}&0\\
\Delta\mathrm{sin}\theta&0&0&-\hbar v_{f}^{G}{k}_{+}	&+\Delta\mathrm{cos}\theta&0\\
0&0&0&0&0&-t_1-t_2-t_3\\
\label{equ:6x6kpmodel}
\end{pmatrix}. 
\end{equation}The physical meaning of those parameters have been explained in the main context.

With the spin degree of freedom, the Hamiltonian of graphene becomes
\begin{equation}
H_{tb}^s=\begin{pmatrix}
H_{tb}& 0 \\
0&H_{tb}\\
\end{pmatrix}
\end{equation} and the basis states expands to $\{|p_z,1,\uparrow\rangle,|p_z,2,\uparrow\rangle,|p_z,3,\uparrow\rangle,|p_z,4,\uparrow\rangle,|p_z,5,\uparrow\rangle,|p_z,6,\uparrow\rangle,|p_z,1,\downarrow\rangle,|p_z,2,\downarrow\rangle,|p_z,3,\downarrow\rangle,|p_z,4,\downarrow\rangle,|p_z,5,\downarrow\rangle,|p_z,6,\downarrow\rangle\}$. Next, a unitary transformation $U$ is applied to the basis states and transforms the Hamiltonian as
\begin{equation}
H_{tb}^s\rightarrow U^{\dagger}H_{tb}^sU. 
\end{equation}
The $12\times 12$ transformation matrix $U$ is chosen to be
\begin{equation}
\begin{pmatrix}
1 & \omega^{-3} & 0 & 0 & 1 & 0 & 0 & 0 & 1 & 1 & 0 & \omega^{3} \\
\omega^{-4} & \omega^{-5} & 0 & 0 & 1 & 0 & 0 & 0 & -1 & \omega^{4} & 0 & \omega^{5} \\
\omega^{4} & \omega^{5} & 0 & 0 & 1 & 0 & 0 & 0 & 1 & \omega^{-4} & 0 & \omega^{-5} \\
1 & \omega^{3} & 0 & 0 & 1 & 0 & 0 & 0 & -1 & 1 & 0 & \omega^{-3} \\
\omega^{-4} & \omega & 0 & 0 & 1 & 0 & 0 & 0 & 1 & \omega^{4} & 0 & \omega^{-1} \\
\omega^{4} & \omega^{-1} & 0 & 0 & 1 & 0 & 0 & 0 & -1 & \omega^{-4} & 0 & \omega \\
0 & 0 & 1 & 1 & 0 & \omega^{-3} & 1 & \omega^{3} & 0 & 0 & 1 & 0 \\
0 & 0 & 1 & \omega^{-4} & 0 & \omega^{-5} & \omega^{4} & \omega^{5} & 0 & 0 & -1 & 0 \\
0 & 0 & 1 & \omega^{4} & 0 & \omega^{5} & \omega^{-4} & \omega^{-5} & 0 & 0 & 1 & 0 \\
0 & 0 & 1 & 1 & 0 & \omega^{3} & 1 & \omega^{-3} & 0 & 0 & -1 & 0 \\
0 & 0 & 1 & \omega^{-4} & 0 & \omega & \omega^{4} & \omega^{-1} & 0 & 0 & 1 & 0 \\
0 & 0 & 1 & \omega^{4} & 0 & \omega^{-1} & \omega^{-4} & \omega & 0 & 0 & -1 & 0 \\
\end{pmatrix}. 
\end{equation}
Expanding the transformed Hamiltonian around $\bm{k}=0$ and excluding the energy levels far away from $\mu^G$ gives
\begin{equation}
\begin{split}
&H_{GG}=\mu^G+\\
&\begin{pmatrix}
	|\frac{3}{2},+1\rangle	&|-\frac{1}{2},+1\rangle	&|\frac{1}{2},+1\rangle	&|-\frac{3}{2},+1\rangle	&|-\frac{3}{2},-1\rangle	&|\frac{1}{2},-1\rangle	&|-\frac{1}{2},-1\rangle	&|\frac{3}{2},-1\rangle	\\
	\hline
	-\Delta\mathrm{cos}\theta	&\hbar {v}_{f}^{G}{k}_{+}	&0	&0	&0	&0	&0	&\Delta\mathrm{sin}\theta	\\
	\hbar v_{f}^{G}{k}_{-}	&+\Delta\mathrm{cos}\theta	&0	&0	&0	&0	&\Delta\mathrm{sin}\theta	&0	\\
	0	&0	&-\Delta\mathrm{cos}\theta	&\hbar v_{f}^{G}{k}_{+}	&0	&\Delta\mathrm{sin}\theta	&0	&0	\\
	0	&0	&\hbar v_{f}^{G}{k}_{-}	&+\Delta\mathrm{cos}\theta	&\Delta\mathrm{sin}\theta	&0	&0	&0	\\
	0	&0	&0	&\Delta\mathrm{sin}\theta	&-\Delta\mathrm{cos}\theta	&-\hbar v_{f}^{G}{k}_{-}	&0	&0	\\
	0	&0	&\Delta\mathrm{sin}\theta	&0	&-\hbar v_{f}^{G}{k}_{+}	&+\Delta\mathrm{cos}\theta	&0	&0	\\
	0	&\Delta\mathrm{sin}\theta	&0	&0	&0	&0	&-\Delta\mathrm{cos}\theta	&-\hbar v_{f}^{G}{k}_{-}	\\
	\Delta\mathrm{sin}\theta	&0	&0	&0	&0	&0	&-\hbar v_f^G{k}_+	&+\Delta\mathrm{cos}\theta	\\
	\end{pmatrix}\\
\end{split}
\label{equ:8x8kpmodel}
\end{equation}
where
\begin{equation}
	\begin{split}
	|\frac{3}{2},+1\rangle =|L_z=1,\eta=+1,\uparrow\rangle \qquad&|-\frac{1}{2},+1\rangle	=|L_z=-1,\eta=+1,\uparrow\rangle\\
	|\frac{1}{2},+1\rangle	=|L_z=1,\eta=+1,\downarrow\rangle\qquad&|-\frac{3}{2},+1\rangle	=|L_z=-1,\eta=+1,\downarrow\rangle\\
	|-\frac{3}{2},-1\rangle	=|L_z=-1,\eta=-1,\downarrow\rangle \qquad&|\frac{1}{2},-1\rangle	=|L_z=1,\eta=-1,\downarrow\rangle\\	
	|-\frac{1}{2},-1\rangle	=|L_z=-1,\eta=-1,\uparrow\rangle \qquad&|\frac{3}{2},-1\rangle =|L_z=1,\eta=-1,\uparrow\rangle\\
	\end{split}
	\label{equ:Jbasis}
\end{equation}
and
\begin{equation}
\begin{pmatrix}
 |L_z=1,\eta=+1\rangle\\
 |L_z=-1,\eta=+1\rangle\\
 |L_z=1,\eta=-1\rangle \\
 |L_z=-1,\eta=-1\rangle\\
\end{pmatrix}=  \frac{1}{\sqrt{6}}
\begin{pmatrix}
  1&\omega^{-4}&\omega^4&1&\omega^{-4}&\omega^{4}\\	
		 \omega^{-3}&\omega^{-5}&\omega^{5}&\omega^{3}&\omega^{1}&\omega^{-1}\\
		\omega^{3}&\omega^{5}&\omega^{-5}&\omega^{-3}&\omega^{-1}&\omega^{1}\\
		 1&\omega^4&\omega^{-4}&1&\omega^{4}&\omega^{-4}\\
\end{pmatrix}
\begin{pmatrix}
|p_z,1\rangle\\
|p_z,2\rangle\\
|p_z,3\rangle\\
|p_z,4\rangle\\
|p_z,5\rangle\\
|p_z,6\rangle\\
\end{pmatrix}
\label{equ:transform}. 
\end{equation}
$J_z,L_z$ are the total angular momentum and orbital angular momentum of $C_3$ rotation in $z$ direction respectively. $C_3|L_z,\eta\rangle=\omega^{-4L_z}|L_z,\eta\rangle$. The original $K_0,K'_0$ being folded onto the $\Gamma$ point gives rise to the additional index $\eta$. The details about the definition of the index $\eta$ are in appendix C. The four states $|L_z=0,\eta=\pm 1,\uparrow(\downarrow)\rangle$ are excluded because their energy levels are much further away from the Fermi level.

The surface states are described by $H_{SS}$:
\begin{equation}
\begin{pmatrix}
\mu^S	&i\hbar v_f^S	k_-\\
-i\hbar v_f^S k_+	&\mu^S	\\
\end{pmatrix}
\end{equation} with the basis states:
\begin{equation}
|\pm\frac{1}{2}\rangle =\sum\limits_{\alpha=Sb,Te}u_{\alpha}|\alpha,p_z,\uparrow(\downarrow)\rangle+v_{\alpha}|\alpha,p_{\pm},\downarrow(\uparrow)\rangle
\label{equ:surfacestates}
\end{equation} where $|p_{\pm}\rangle=\mp\frac{1}{\sqrt{2}}(|p_x\rangle\pm i|p_y\rangle)$ and $u_{\alpha},\,v_{\alpha}$ are assumed to be real.

The hybridization between $p$ orbitals can be decoupled to two types ($V_{pp\pi}$,$V_{pp\sigma}$). In this case, the hybridization between $p_z$ orbital of carbon atoms and $p_z$ / $p_{\pm}$ orbital from $\alpha$ ($\alpha$=Te,Sb) is proportional to $V^{\alpha}_{p_z}$ / $V^{\alpha}_{p_{\pm}}$ and we have the following decompositions:
\begin{equation}
\begin{split}
V^{\alpha}_{p_z}&=V^{\alpha}_{\perp}=V^{\alpha}_{pp\pi}\text{sin}^2\theta^{\alpha}+V^{\alpha}_{pp\sigma}\text{cos}^2\theta^{\alpha}\\
V^{\alpha}_{p_{\pm}}&=\mp V^{\alpha}_{\parallel}e^{\pm i\varphi^{\alpha}}=\mp\frac{1}{\sqrt{2}}(V^{\alpha}_{pp\sigma}-V^{\alpha}_{pp\pi})\text{cos}\theta^{\alpha}\text{sin}\theta^{\alpha}e^{\pm i\varphi^{\alpha}}\\
\end{split}
\end{equation}
where $\theta^{\alpha}$ stands for the angle between $z$ axis and $\mathcal{L}^{\alpha}$, $\varphi^{\alpha}$ denotes the angle between the positive $x$ axis and the projection of $\mathcal{L}^{\alpha}$ on the $x-y$ plane. Here $\mathcal{L}^{\alpha}$ denotes the vector connecting carbon atom and $\alpha$ atom and pointing to the carbon atom.
For the hybridization submatrix $H_{GS}$,  the sixteen terms have been deduced from real space tight-binding method (up to the order of $\bm{k}$):
\begin{equation}
\begin{split}
-\langle -\frac{3}{2},-1|H|-\frac{1}{2}\rangle ^*=\langle \frac{3}{2},+1|H|\frac{1}{2}\rangle &= \frac{\sqrt{2}}{4} V^{Sb}_{\perp}u_{Sb}|\bm{a_1}|k_-\\
\langle -\frac{3}{2},-1|H|\frac{1}{2}\rangle^*=\langle \frac{3}{2},+1|H|-\frac{1}{2}\rangle &= 0\\
-\langle \frac{1}{2},-1|H|-\frac{1}{2}\rangle ^*=\langle -\frac{1}{2},+1|H|\frac{1}{2}\rangle &=  \frac{\sqrt{6}}{4} V^{Sb}_{\perp}u_{Sb}|\bm{a_1}|k_+\\
-\langle \frac{1}{2},-1|H|\frac{1}{2}\rangle^*=\langle -\frac{1}{2},+1|H|-\frac{1}{2}\rangle &=-\sqrt{6}iV^{Te}_{\parallel}v_{Te}+\frac{\sqrt{6}i}{2}V^{Sb}_{\parallel}v_{Sb}\\
-\langle -\frac{1}{2},-1|H|-\frac{1}{2}\rangle^*=\langle \frac{1}{2},+1|H|\frac{1}{2}\rangle &= -\frac{3\sqrt{2}i}{2} V^{Sb}_{\parallel}v_{Sb}\\
-\langle -\frac{1}{2},-1|H|\frac{1}{2}\rangle ^*=\langle \frac{1}{2},+1|H|-\frac{1}{2}\rangle &= \frac{\sqrt{2}}{4} V^{Sb}_{\perp}u_{Sb}|\bm{a_1}|k_-\\
\langle \frac{3}{2},-1|H|-\frac{1}{2}\rangle ^*=\langle -\frac{3}{2},+1|H|\frac{1}{2}\rangle &= -\frac{\sqrt{2}i}{2} V^{Sb}_{\parallel}v_{Sb}|\bm{a_1}|k_-\\
-\langle \frac{3}{2},-1|H|\frac{1}{2}\rangle ^*=\langle -\frac{3}{2},+1|H|-\frac{1}{2}\rangle &= \frac{\sqrt{6}}{4} V^{Sb}_{\perp}u_{Sb}|\bm{a_1}|k_+\\
\end{split}
\end{equation}
where $\langle\psi_1'|H(\bm{k})|\psi_2'\rangle=\langle\psi_1|T^{-1}H(\bm{k})T|\psi_2\rangle=\langle\psi_1|H^*(-\bm{k})|\psi_2\rangle$ is used.
Two examples of the above calculations are
\begin{equation}
\begin{split}
\langle \frac{3}{2},+1|H|\frac{1}{2}\rangle&=\langle L_z=1,\eta=+1,\uparrow|H|\frac{1}{2}\rangle\\
&=\sum\limits_{\alpha}u_{\alpha}\langle L_z=1,\eta=+1,\uparrow|H|\alpha,p_z,\uparrow\rangle\\
&=\sum\limits_{\alpha,n}u_{\alpha}c_n^*\langle p_z,n,\uparrow|H|\alpha,p_z,\uparrow\rangle\\
&=u_{Te}V_{\perp}^{Te}(1+\omega^4+\omega^{-4}+1+\omega^4+\omega^{-4})/\sqrt{6}\\
&\quad+u_{Sb}V_{\perp}^{Sb}[(1+\omega^4)e^{-i\bm{k}\cdot\bm{a_1}}+(\omega^{-4}+1)e^{-i\bm{k}\cdot\bm{a_2}}+(\omega^4+\omega^{-4})]/\sqrt{6}\\
&=\frac{\sqrt{2}}{4}u_{Sb}V^{Sb}_{\perp}|\bm{a_1}|k_-
\end{split}
\end{equation}
and
\begin{equation}
\begin{split}
\langle \frac{3}{2},+1|H|-\frac{1}{2}\rangle&=\langle L_z=1,\eta=+1,\uparrow|H|-\frac{1}{2}\rangle\\
&=\sum\limits_{\alpha}v_{\alpha}\langle L_z=1,\eta=+1,\uparrow|H|\alpha,p_-,\uparrow\rangle\\
&=\sum\limits_{\alpha,n}u_{\alpha}c_n^*\langle p_z,n,\uparrow|H|\alpha,p_-,\uparrow\rangle\\
&=v_{Te}V_{\parallel}^{Te}(1\cdot\omega^6+\omega^4\cdot\omega^4+\omega^{-4}\cdot\omega^2+1+\omega^4\cdot\omega^{-2}+\omega^{-4}\cdot\omega^{-4})/\sqrt{6}\\
&\quad+v_{Sb}V_{\parallel}^{Sb}[(1\cdot\omega^{-2}+\omega^4)e^{-i\bm{k}\cdot\bm{a_1}}+(\omega^{-4}\cdot\omega^{-6}+1\cdot\omega^{-4})e^{-i\bm{k}\cdot\bm{a_2}}\\
&\quad+(\omega^4\cdot\omega^2+\omega^{-4}\cdot\omega^4)]/\sqrt{6}\\
&=0
\end{split}
\end{equation}
where the hybridization is limited between states with the same spin and the transformation $|\frac{3}{2},+1\rangle=\sum\limits_{n}c_n|p_z,n,\uparrow\rangle$ in Eq. [\ref{equ:Jbasis}] and Eq. [\ref{equ:transform}] is used. The complete Hamiltonian $H_{GS}$ for the coupling between graphene and topological surface states of Sb$_2$Te$_3$ films is written as
\begin{equation}
\begin{pmatrix}[1.5]
\frac{\sqrt{2}}{4} V^{Sb}_{\perp}u_{Sb}|\bm{a_1}|k_-&0\\
 \frac{\sqrt{6}}{4} V^{Sb}_{\perp}u_{Sb}|\bm{a_1}|k_+&-\sqrt{6}iV^{Te}_{\parallel}v_{Te}+\frac{\sqrt{6}i}{2}V^{Sb}_{\parallel}v_{Sb}\\
-\frac{3\sqrt{2}i}{2} V^{Sb}_{\parallel}v_{Sb}& \frac{\sqrt{2}}{4} V^{Sb}_{\perp}u_{Sb}|\bm{a_1}|k_-\\
-\frac{\sqrt{2}i}{2} V^{Sb}_{\parallel}v_{Sb}|\bm{a_1}|k_-& \frac{\sqrt{6}}{4} V^{Sb}_{\perp}u_{Sb}|\bm{a_1}|k_+\\
0&-\frac{\sqrt{2}}{4} V^{Sb}_{\perp}u_{Sb}|\bm{a_1}|k_+\\
-\sqrt{6}iV^{Te}_{\parallel}v_{Te}+\frac{\sqrt{6}i}{2}V^{Sb}_{\parallel}v_{Sb}& -\frac{\sqrt{6}}{4} V^{Sb}_{\perp}u_{Sb}|\bm{a_1}|k_-\\
-\frac{\sqrt{2}}{4} V^{Sb}_{\perp}u_{Sb}|\bm{a_1}|k_+&-\frac{3\sqrt{2}i}{2} V^{Sb}_{\parallel}v_{Sb}\\
- \frac{\sqrt{6}}{4} V^{Sb}_{\perp}u_{Sb}|\bm{a_1}|k_-&\frac{\sqrt{2}i}{2} V^{Sb}_{\parallel}v_{Sb}|\bm{a_1}|k_+\\
\end{pmatrix}. 
\label{equ:HGS}
\end{equation}

Now all the terms in $H_{GG}$, $H_{SS}$ and $H_{GS}$ have been explicitly shown and the parameters in them can be determined by fitting the the \emph{ab initio} band structure, as listed in the Table \ref{tab:fittingpara}. 

\begin{table}[H]
\centering
\begin{tabular}{|c|c|}
\hline
$\mu^G$ (eV) & 0.008 \\
\hline
$\mu^S$ (eV) & -0.041 \\
\hline
$v_f^G$ (m$\cdot$s$^{-1}$) & $0.777\times 10^6$ \\
\hline
$v_f^S $ (m$\cdot$s$^{-1}$) & $0.278\times 10^6$ \\
\hline
$\Delta$ (eV) & 0.010 \\
\hline
$\theta$ & 2.904 \\
\hline
$V_{\perp}^{Sb}u_{Sb}$ (eV) & -0.141 \\
\hline
$V_{\parallel}^{Te}v_{Te}$ (eV) & 0.014 \\
\hline
$V_{\parallel}^{Sb}v_{Sb}$ (eV) & -0.005 \\
\hline
\end{tabular}
\caption{Fitting parameters of the effective Hamiltonian $H_{full}$ in Eq. [\ref{equ:equ1}] for the \emph{ab initio} band structure in Fig. [\ref{Fig:fig1}].}
\label{tab:fittingpara}
\end{table}

\renewcommand{\theequation}{C.\arabic{equation}}
  \renewcommand{\thetable}{C.\arabic{table}}
\setcounter{equation}{0}  
\setcounter{table}{0}
\subsection{Definition of the index $\eta$}
For a pristine graphene with atom 1 and 2 in the primitive cell in Fig. \ref{Fig:fig1}(b), the eigenstates of the Hamiltonian at the Dirac point ($|K\rangle$ and $|K'\rangle$) should take the form of
\begin{equation}
\begin{pmatrix}
c_1\\
c_2\\
\end{pmatrix}
\end{equation}
with the basis being $\{|p_z,1\rangle, |p_z,2\rangle\}$. If we enlarge the basis set to $\{|p_z,1\rangle, |p_z,2\rangle,|p_z,3\rangle, |p_z,4\rangle,|p_z,5\rangle, |p_z,6\rangle\}$, according to Bloch theorem, the eigenstates at $k=K,K'$ become
\begin{equation}
\begin{pmatrix}
c_1\\
c_2\\
c_1\omega^4\\
c_2\omega^{-4}\\
c_1\omega^{-4}\\
c_2\omega^4\\
\end{pmatrix}\rightarrow K,\quad
\begin{pmatrix}
c_2\omega^4\\
c_1\\
c_2\\
c_1\omega^4\\
c_2\omega^{-4}\\
c_1\omega^{-4}\\
\end{pmatrix}\rightarrow K'
\label{equ:Kc1c2}
\end{equation}
where $\omega=exp(\frac{i\pi}{6})$. These two sates be represented by $|K,c_1,c_2\rangle$ and $|K',c_1,c_2\rangle$ respectively and $|K',c_1,c_2\rangle=C_6|K,c_1,c_2\rangle$. More importantly, we notice that
\begin{equation}
\begin{split}
C_3|K,1,0\rangle=\omega^{-4}|K,1,0\rangle,\quad &C_3|K',1,0\rangle=\omega^{-4}|K',1,0\rangle\\
C_3|K,0,1\rangle=\omega^{4}|K,0,1\rangle,\quad &C_3|K',0,1\rangle=\omega^{4}|K',0,1\rangle\\
\end{split}
\end{equation}
The states $|K(K'),1,0\rangle$ are eigenstates of $C_3$ rotation operator with the orbital angular momentum in $z$ direction $L_z$ being $+1$, while $|K(K'),0,1\rangle$ correspond to $L_z$ being $-1$. According to the definitions of $|K(K'),c_1,c_2\rangle$ in Eq. [\ref{equ:Kc1c2}], $|K(K'),1,0\rangle$ are projections of $|K(K')\rangle$ on the sublattice containing atom 1 and $|K(K'),0,1\rangle$ are projections of $|K(K')\rangle$ on the sublattice containing atom 2. To classify the doubly degenerate states at $K,K'$ further, we introduce $C_6$, the six-fold rotation operator. And we find
\begin{equation}
\begin{split}
C_6(|K,1,0\rangle+\omega^{-4}|K',1,0\rangle)&=\omega^4(|K,1,0\rangle+\omega^{-4}|K',1,0\rangle)=\omega^4|C_6,L_z'=-2\rangle\\
C_6(\omega^3|K,1,0\rangle+\omega^{5}|K',1,0\rangle)&=\omega^{-2}(\omega^3|K,1,0\rangle+\omega^{5}|K',1,0\rangle)=\omega^{-2}|C_6,L_z'=1\rangle\\
\end{split}
\end{equation}
In the graphene$/$Sb$_2$Te$_3$ hetero-junction, the $C_3$ rotation symmetry is preserved but $C_6$ rotation symmetry is broken. So to take advantage of the classifications above, we introduce an additional index $\eta$ and set
\begin{equation}
|L_z=1,\eta=+1\rangle=|C_6,L_z'=-2\rangle,\quad |L_z=1,\eta=-1\rangle=|C_6,L_z'=1\rangle
\label{equ:eta1}
\end{equation}
The situation of the states $|K(K'),0,1\rangle$ is similar.
\begin{equation}
\begin{split}
C_6(\omega^{-5}|K,0,1\rangle+\omega^{5}|K',0,1\rangle)&=\omega^{2}(\omega^{-5}|K,0,1\rangle+\omega^{5}|K',0,1\rangle)=\omega^{2}|C_6,L_z'=-1\rangle\\
C_6(\omega^4|K,0,1\rangle+\omega^{-4}|K',0,1\rangle)&=\omega^{-4}(\omega^4|K,0,1\rangle+\omega^{-4}|K',0,1\rangle)=\omega^{-4}|C_6,L_z'=2\rangle\\
|L_z=-1,\eta=+1\rangle=|C_6,&L_z'=-1\rangle,\quad |L_z=-1,\eta=-1\rangle=|C_6,L_z'=2\rangle\\
\end{split}
\label{equ:eta2}
\end{equation}
It is easy to check that the definitions of $|L_z,\eta\rangle$ here in Eq. [\ref{equ:eta1}] and Eq. [\ref{equ:eta2}] are consistent with those in Eq. [\ref{equ:transform}]. To sum up, the electron state $|L_z,\eta\rangle$ is an eigenstate of $C_6$ rotation operator and also a linear supercomposition of $|K\rangle$ projected on one sublattice and $|K'\rangle$ projected on the other sublattice. According to the definitions above, the time-reversal operation changes the sign of both $L_z$ and $\eta$.

\renewcommand{\theequation}{D.\arabic{equation}}
  \renewcommand{\thetable}{D.\arabic{table}}
\setcounter{equation}{0}  
\setcounter{table}{0}
\subsection{Possible coupling processes between $| -\frac{1}{2},-1\rangle$ and $| \frac{1}{2},+1\rangle$}
The spin-flip coupling process should include the coupling between these graphene states with the topological surface states $|\pm\frac{1}{2}\rangle$, which can be decomposed into $p$ orbitals of Sb and Te atoms as in Eq. [\ref{equ:surfacestates}]. Bearing the decompositions in mind, we have the following perturbative terms that produce linear momentum dependence (the energy difference in the denominator is assumed to be constant and ignored)
\begin{equation}
\begin{split}
\langle \frac{1}{2},1|H|\alpha,p_+,\downarrow\rangle \underline{\langle \alpha,p_+,\downarrow|H|-\frac{1}{2},-1\rangle} &\propto k_-\\
\underline{\langle \frac{1}{2},1|H|\alpha,p_z,\uparrow\rangle}{\langle \alpha,p_z,\uparrow|H|-\frac{1}{2},-1\rangle} &\propto k_-\\
\underline{\langle \frac{1}{2},1|H|\alpha,p_-,\uparrow\rangle} \langle \alpha,p_-,\uparrow|H|-\frac{1}{2},-1\rangle &\propto k_-\\
{\langle \frac{1}{2},1|H|\alpha,p_z,\downarrow\rangle}\underline{\langle \alpha,p_z,\downarrow|H|-\frac{1}{2},-1\rangle} &\propto k_-\\
\end{split}
\end{equation}
with $\alpha=Te,Sb$ for TIs. Here the first two terms are from $\langle\frac{1}{2},1|H|J^S_z=\frac{1}{2}\rangle\langle J^S_z=\frac{1}{2}|H|-\frac{1}{2},1\rangle$ and the last two are from $\langle\frac{1}{2},1|H|J^S_z=-\frac{1}{2}\rangle\langle J^S_z=-\frac{1}{2}|H|-\frac{1}{2},1\rangle$. The underlined terms, proportional to $k_-$, involve the spin-flip hopping between graphene and TI. 
Therefore, only second order (or higher order) perturbation processes can give rise to the linear momentum dependence and that is why the Fermi velocity of the resulting state is much smaller than that of the TI surface states.

\renewcommand{\theequation}{E.\arabic{equation}}
  \renewcommand{\thetable}{E.\arabic{table}}
\setcounter{equation}{0}  
\setcounter{table}{0}
\subsection{Mean field theory for ``heavy'' Dirac fermions with interaction}
Based on the third order L$\ddot{o}$wdin perturbation theory \cite{winkler2003spin}, we obtain the effective Hamiltonian near the $\Gamma$ point
\begin{equation}
\begin{split}
H_{D}=&\begin{pmatrix}
\tilde{C}_1k_x k_y^2-C_1k_x^3 & iC_2 k^2 k_- + iC_3k_-\\
-iC_2k^2 k_+ -iC_3k_+ &- \tilde{C}_1 k_x k_y^2+C_1k_x^3\\
\end{pmatrix} \\
&+(e_0-C_0k^2)I_{2\times2}\\
=&(C_2k^2+C_3)\hat{z}\cdot({\bf \sigma}\times{\bf k})-\frac{C_1}{2}(k_+^3+k_-^3)\sigma_z\\
&+(e_0-C_0k^2)I_{2\times2}\\
\end{split}
\label{Eq:Heff}
\end{equation}
where $C_0=116.1186$ eV, $\tilde{C}_1=96.7254$ eV, $C_1=32.2418$ eV, $C_2=936.4909$ eV, $C_3=0.02039$ eV. In the $H_{eff}$, we have set lattice constant $a=1$ and $k_{x,y}$ is dimensionless. We have made use of the fact that $\tilde{C}_1=3C_1$, and found a term $-\frac{C_1}{2}(k_+^3+k_-^3)\sigma_z$ to be exactly the hexagon wrapping term in Sb$_2$Te$_3$. $e_0$ is changed to be zero for simplicity. $H_{eff}$ indeed captures the correct physics in the vicinity of $\Gamma$ point. The higher order perturbation included, the better $H_{D}$ performs. In our case, it is reasonable to include perturbations up to the third order.

Before considering the interaction effects, it is advisable to first look at the possible mass terms to $H_{D}$. Consider a general mass term $M=\sum_{i\in(x,y,z)} m_i\sigma_i$,
\begin{equation}
\begin{split}
H_{D}=&((C_2k^2+C_3)k_y+m_x)\sigma_x+(m_y-(C_2k^2+C_3)k_x)\sigma_y)\\
&+(\tilde{C}_1k_x k_y^2-C_1k_x^3+m_z)\sigma_z  \\
E_k=&\pm[((C_2k^2+C_3)k_y+m_x)^2+(m_y-(C_2k^2+C_3)k_x)^2  \\
&+(\tilde{C}_1k_x k_y^2-C_1k_x^3+m_z)^2]^{\frac{1}{2}}\\
\end{split}
\end{equation} where the identity term $C_0k^2I_{2\times2}$ is ignored. When $m_x=m_y=0$, $m_z\sigma_z$ is obviously a mass term. When $m_y=m_z=0$, being gapless requires $k_x=0$ and the energy dispersion is 
\begin{equation}
E_g^x=2[(C_2k_y^2+C_3)k_y+m_x]
\end{equation}
So we could always find a possible $k_y$, where $E_g^x=0$ when $k_x=0$. When $m_x=m_z=0$, being gapless requires $k_y=0$ and the corresponding eigen energy becomes
\begin{equation}
E_g^y=2\sqrt{(m_y-(C_2k_x^2+C_3)k_x)^2+(C_1k_x^3)^2}. 
\end{equation}
In this case, we cannot find a gapless point in the band structure. Thus, the general form for the mass terms should be $M=m_y\sigma_y+m_z\sigma_z$.

We consider an onsite Hubbard repulsion,
\begin{equation}
H=\sum_{\langle r,r'\rangle}\Psi^{\dagger}(r) H_0(r,r') \Psi(r')+ U\sum_{r} \psi_1^{\dagger}(r) \psi_1(r) \psi_2^{\dagger}(r) \psi_2(r)
\end{equation}
Here we have defined real space basis $\Psi^{\dagger}(r)=(\psi_1^{\dagger}(r),\psi_2^{\dagger}(r))$, and a real space Hamiltonian $H(r,r')=H(r-r')$ related to our effective Hamiltonian via Fourier transformation
\begin{equation}
H_0=\sum_{\langle r,r'\rangle}c_r^{\dagger}H(r,r')c_{r'}=\sum_k c_k^{\dagger}c_k H_{eff}(k). 
\end{equation}

Based on previous mass terms analysis, we expect interaction-induced mass terms to be proportional to $\sigma_y$ and $\sigma_z$. Such mass terms will naturally arises if we consider the following decomposition scheme to the four-fermion interaction:
\begin{eqnarray}
\psi_1^{\dagger}\psi_1\psi_2^{\dagger}\psi_2&=&(\psi_1^{\dagger}\psi_1)(\psi_2^{\dagger}\psi_2)-(\psi_1^{\dagger}\psi_2)(\psi_2^{\dagger}\psi_1) \nonumber \\
&=&\textcircled{1}-\textcircled{2}. 
\end{eqnarray}
We define the following order parameters:
\begin{eqnarray}
\delta_0&=&\langle\psi_1^{\dagger}\psi_2\rangle=\delta e^{i\theta} \nonumber \\
\delta_0^*&=&\langle\psi_2^{\dagger}\psi_1\rangle=\delta e^{-i\theta} \nonumber \\
\delta_1&=&\langle\psi_1^{\dagger}\psi_1\rangle \nonumber \\
\delta_2&=&\langle\psi_2^{\dagger}\psi_2\rangle, 
\end{eqnarray}
and we could perform the mean field theory based on the above order parameters,
\begin{eqnarray}
\textcircled{1}&=&(\psi_1^{\dagger}\psi_1-\delta_1+\delta_1)(\psi_2^{\dagger}\psi_2-\delta_2+\delta_2) \nonumber \\
&=&-\delta_1\delta_2+\psi_1^{\dagger}\psi_1\delta_2+\psi_2^{\dagger}\psi_2\delta_1 \nonumber \\
&=&-\delta_1\delta_2+\Psi^{\dagger}
\begin{pmatrix}
\delta_2 & 0 \\
0 & \delta_1
\end{pmatrix}\Psi \nonumber \\
&=&A^{-2}\frac{m_z^2-m_0^2}{4}+A^{-1}\Psi^{\dagger}
\begin{pmatrix}
\frac{m_0-m_z}{2} & 0 \\
0 & \frac{m_z+m_0}{2}
\end{pmatrix} \Psi \nonumber \\
&=&(A^{-2}\frac{m_z^2}{4}-A^{-1}\frac{m_z\Psi^{\dagger}\sigma_z\Psi}{2})-(A^{-2}\frac{m_0^2}{4}-A^{-1}\frac{m_0\Psi^{\dagger}\sigma_0\Psi}{2}). 
\end{eqnarray}
Here we have introduced the area of the system $A$ to make order parameters dimensionless.
\begin{eqnarray}
A^{-1}m_z&=&\delta_1-\delta_2=\langle\Psi^{\dagger}\sigma_z\Psi\rangle \nonumber \\
A^{-1}m_0&=&\delta_1+\delta_2=\langle\Psi^{\dagger}\sigma_0\Psi\rangle. 
\end{eqnarray}
Similarly,
\begin{eqnarray}
\textcircled{2}&=&(\psi_1^{\dagger}\psi_2-\delta_0+\delta_0)(\psi_2^{\dagger}\psi_1-\delta_0^*+\delta_0^*) \nonumber \\
&=&-|\delta_0|^2+\psi_1^{\dagger}\psi_2\delta_0^*+\psi_2^{\dagger}\psi_1\delta_0 \nonumber \\
&=&-|\delta_0|^2+\Psi^{\dagger}
\begin{pmatrix}
0 & \delta_0^* \\
\delta_0 & 0
\end{pmatrix} \Psi \nonumber \\
&=&-|\delta_0|^2+\Psi^{\dagger} (\delta_0\cos\theta \sigma_x + \delta_0\sin\theta \sigma_y) \Psi \nonumber \\
&=&-A^{-2}\frac{m_x^2+m_y^2}{4}+A^{-1}\Psi^{\dagger}(\frac{m_x}{2}\sigma_x+\frac{m_y}{2}\sigma_y) \Psi,  \nonumber \\
\end{eqnarray}
where we have defined
\begin{eqnarray}
A^{-1}m_x&=&\delta_0+\delta_0^*=\langle\psi_1^{\dagger}\psi_2+\psi_2^{\dagger}\psi_1\rangle=\langle\Psi^{\dagger}\sigma_x\Psi\rangle \nonumber \\
A^{-1}m_y&=&\frac{\delta_0-\delta_0^*}{i}=\langle-i\psi_1^{\dagger}\psi_2+i\psi_2^{\dagger}\psi_1\rangle=\langle\Psi^{\dagger}\sigma_y\Psi\rangle. 
\end{eqnarray}
In the previous section, we have shown that only mass terms that couples to either $\sigma_z$ or $\sigma_y$ will gap the system. So we will consider the contributions from order $m_z$ and $m_y$. In the mean field approximation, we write down the Hamiltonian and Fourier transform it into momentum space,
\begin{eqnarray}
H&=&\sum_k \Psi^{\dagger}(k)H_{eff}(k) \Psi(k) + U \sum_r [\textcircled{1}-\textcircled{2}] \nonumber \\
&=& \sum_k \Psi^{\dagger}(k)H_{eff}(k) \Psi(k)- \frac{U}{2A}\sum_r \Psi^{\dagger}(r)(m_y\sigma_y+m_z\sigma_z)\Psi(r) \nonumber \\
&&+\frac{U}{4A} (m_y^2+m_z^2) \nonumber \\
&=&\sum_k \Psi^{\dagger}(k)[H_{eff}(k)-\frac{U}{2A}(m_y\sigma_y+m_z\sigma_z)]\Psi(k)+\frac{U}{4A}(m_y^2+m_z^2) \nonumber \\
&=&\sum_k \Psi^{\dagger}(k)[H_{eff}(k)-\frac{U}{2}(m_y\sigma_y+m_z\sigma_z)]\Psi(k)+\frac{U}{4}(m_y^2+m_z^2). 
\end{eqnarray}

Here we have redefined $U\rightarrow U\times A$ and the new $U$ has the dimension of energy. The above mass terms obviously both break time reversal symmetry, and thus gap the surface states. A difference between the two mass terms is that $m_y\sigma_y$ breaks three fold rotation symmetry $C_3$, while $m_z\sigma_z$ preserves $C_3$ symmetry. Therefore, $m_y\sigma_y$ leads to a nematic phase. If the system is in this nematic regime, such nematic ordering could be detected via scanning tunneling microscopy (STM) measurement. In general, both terms here will gap the surface states and result in the spontaneous half quantum hall state on the surface.

\renewcommand{\theequation}{F.\arabic{equation}}
  \renewcommand{\thetable}{F.\arabic{table}}
\setcounter{equation}{0}  
\setcounter{table}{0}
\subsection{Critical interaction solution for self-consistent equations}

The zero temperature free energy $F$ is
\begin{eqnarray}
F&=&\frac{U}{4}(m_y^2+m_z^2)-\sum_k \sqrt{d_x(k)^2+(d_y(k)-\frac{U}{2}m_y)^2+(d_z(k)-\frac{U}{2}m_z)^2} \nonumber \\
&=&\frac{1}{U}(\Delta_y^2+\Delta_z^2)-\sum_k \sqrt{d_x(k)^2+(d_y(k)-\Delta_y)^2+(d_z(k)-\Delta_z)^2}, 
\end{eqnarray}
where $\Delta_{y,z}=\frac{U}{2}m_{y,z}$. Self-consistent equations for order parameter $\Delta_i$ $(i=y,z)$ can be obtained by minimizing $F$:
\begin{eqnarray}
0&=&\frac{\partial F}{\partial \Delta_i} \nonumber \\
&=&\frac{2}{U}\Delta_i-\sum_k \frac{\Delta_i-d_i}{\sqrt{d_x(k)^2+(d_y(k)-\Delta_y)^2+(d_z(k)-\Delta_z)^2}}
\end{eqnarray}

In two dimensions, the density of states (DOS) for a linear dispersion at the Dirac point is vanishing, but DOS for a cubic dispersion is diverging. Thus, in the presence of interactions, systems with a linear dispersion require finite interaction strength to develop ordering, while systems with a cubic dispersion will develop instabilities. In our system, there is an interesting competition between linear terms and cubic terms. When linear term is vanishingly small, we expect the system will spontaneously develop TR breaking ordering ($m_y$ and $m_z$). As we increase the linear term, the critical interaction required to develop the above order parameters also increases. To verify our expectation, we aim at finding a relation between critical interaction $U_c$ and linear coefficient $C_3$, while leaving the cubic coefficient $C_2$ fixed. Recall that the critical interaction happens when the limit $\Delta_i\rightarrow 0$ takes place.
\begin{eqnarray}
\frac{1}{U_c}&=&\frac{1}{2}\lim_{\Delta_{y,z}\rightarrow 0} \sum_k \frac{1}{\sqrt{d_x(k)^2+(d_y(k)-\Delta_y)^2+(d_z(k)-\Delta_z)^2}} \nonumber \\
&&-\frac{1}{2}\lim_{\Delta_{y,z}\rightarrow 0} \frac{\sum_k \frac{d_i(k)}{\sqrt{d_x(k)^2+(d_y(k)-\Delta_y)^2+(d_z(k)-\Delta_z)^2}}}{\Delta_i}
\end{eqnarray}
In the above equation, the first limit is easy to evaluate. For the second limit, the denominator $\Delta_i\rightarrow 0$, and meanwhile, the numerator also approaches zero under this limit, because $\frac{d_i(k)}{\sqrt{d_x(k)^2+(d_y(k)-\Delta_y)^2+(d_z(k)-\Delta_z)^2}}$ is an odd function in $k_i$. Therefore, this limit is a ``$\frac{0}{0}$" type limit, which can be evaluated with the help of the L'Hospital's rule:
\begin{eqnarray}
&&\lim_{\Delta_{y,z}\rightarrow 0} \frac{\sum_k \frac{d_i(k)}{\sqrt{d_x(k)^2+(d_y(k)-\Delta_y)^2+(d_z(k)-\Delta_z)^2}}}{\Delta_i} \nonumber \\
&&=\lim_{\Delta_{y,z}\rightarrow 0} \frac{\sum_k \frac{d}{d \Delta_i} [\frac{d_i(k)}{\sqrt{d_x(k)^2+(d_y(k)-\Delta_y)^2+(d_z(k)-\Delta_z)^2}}]}{\frac{d}{d \Delta_i} \Delta_i} \nonumber \\
&&=\lim_{\Delta_{y,z}\rightarrow 0} \sum_k \frac{d_i(k)(d_i(k)-\Delta_i)}{[d_x(k)^2+(d_y(k)-\Delta_y)^2+(d_z(k)-\Delta_z)^2]^{\frac{3}{2}}} \nonumber \\
&&=\sum_k \frac{d_i(k)^2}{[d_x(k)^2+d_y(k)^2+d_z(k)^2]^{\frac{3}{2}}}. 
\end{eqnarray}
Finally, we arrived at the analytical expression of critical interaction strength $U_c$:
\begin{equation}
\frac{1}{U_c}=\frac{1}{2}\sum_k [\frac{1}{\sqrt{d_x(k)^2+d_y(k)^2+d_z(k)^2}}-\frac{d_i(k)^2}{[d_x(k)^2+d_y(k)^2+d_z(k)^2]^{\frac{3}{2}}}]. 
\end{equation}
In the continuum limit,
\begin{equation}
\frac{1}{U_c A}=\frac{1}{2}\int \frac{d^2k}{(2\pi)^2} [\frac{1}{\sqrt{d_x(k)^2+d_y(k)^2+d_z(k)^2}}-\frac{d_i(k)^2}{[d_x(k)^2+d_y(k)^2+d_z(k)^2]^{\frac{3}{2}}}], 
\end{equation}
where $A$ is the area of the sample and the calculation of the k-space integral involves a choice of momentum cut-off $\Lambda$.


\end{document}